\documentclass[final]{svjour3}
\usepackage{graphicx}
\usepackage{amssymb}
\usepackage{mathptmx}
\journalname{Journal of Low Temperature Physics}

\begin{document}

\newcommand{\hdblarrow}{H\makebox[0.9ex][l]{$\downdownarrows$}-}
\title{Twenty years of 
magnon Bose condensation and spin current superfluidity  in
$^3$He-B}

\author{G.E. Volovik}

\institute{
G.E. Volovik \at
Low Temperature Laboratory, 
Helsinki University of Technology,
P.O.Box 5100, FIN-02015 HUT, Finland, and
L.D. Landau Institute for Theoretical Physics, 119334
 Moscow, Russia\\ Tel.: +358-9-4512963\\ Fax: +358-9-4512969\\
\email{volovik@boojum.hut.fi}}

\date{Received: 02 April 2008 / Accepted: 16 July 2008}

\maketitle

\begin{abstract}
20 years ago a new quantum state of matter was discovered and 
identified  \cite{1,2,3,4,5,6,7}. The observed dynamic  quantum
state of spin precession in superfluid $^3$He-B  bears the
properties of spin current superfluidity, Bose condensation of 
spin waves -- magnons,  off-diagonal long-range order and related
phenomena of quantum coherence.
\keywords{
superfluid $^3$He, spin-current superfluidity, magnon Bose-Einstein
condensation
}
\PACS{67.30.er \and 75.30.Ds \and 67.10.Ba}
\end{abstract}

\section{Introduction}

Nature knows different types of ordered states. 

One major class is represented  by equilibrium macroscopic ordered
sta\-tes exhibiting spontaneous  breaking of symmetry.  This class
contains crystals; nematic, cholesteric and other liquid
crystals;  different types of ordered magnets
(antiferromagnets,  ferromagnets, etc.); superfluids,
superconductors and Bose condensates; all types of Higgs fields
in high energy physics; etc. The important sub\-class\-es of this
class contain  systems with ma\-croscopic quantum coherence
exhibiting off-diagonal long-range order (ODLRO),  and/or
nondissipative superfluid currents (mass current,
spin current, electric current, hypercharge current, etc.). The
class of ordered systems is characterized by rigidity, stable
gradients of  order parameter (non-dissipative currents in quantum
coherent systems), and  topologically stable defects (vortices,
solitons, cosmic strings, monopoles, etc.).

A second large class is presented by  dynamical
systems out of equilibrium. Ordered states may emerge under
external flux of energy. Examples are the coherent emission from
lasers; water flow in a draining bathtub; pattern formation in
dissipative systems; etc. 

Some of the latter dynamic systems can be close   to stationary
equilibrium systems of the first class. For example, ultra-cold
gases in optical traps are not fully equilibrium states since
the number of atoms in the trap is not conserved, and thus the
steady state requires pumping. However, if the decay is small
then the system  is close to an equilibrium Bose
condensate, and experiences all the corresponding superfluid
properties. Bose condensation of quasiparticles  whose
number is not conserved is a timely topic in present
literature: this is Bose condensation of magnons, rotons,
phonons, polaritons, excitons, etc.

There are two different schools in the study of the  Bose
condensation of quasiparticles.  In one of them, Bose
condensation  (and  ODLRO) of quasiparticles is used for 
describing an equilibrium state with diagonal
long-range order, such as crystals, and magnets (see \cite{Radu,Giamarchi}
and references therein). This is somewhat contradictory, since the
essentially non-equilibrium phenomenon of condensation of the
non-conserved  quasiparticles cannot be used for the
description of a true equilibrium state (see e.g. \cite{Mills} and  Appendix H).
Actually  Bose condensation here serves as the instrument for the
desciption of the initial stage of the soft instability which
leads to symmetry breaking and formation of the true equilibrium
ordered state of the first class.  For example, using the 
language of  Bose condensation of phonons one can describe the
soft mechanism of formation of solid crystals
\cite{Kohn}. In the same way Bose condensation   of magnons
can be used for the description of the soft mechanism of
formation of ferromagnetic and antiferromagnetic states (see
e.g.
\cite{Nikuni}). 
The growth of a single mode in the non-linear process after 
a hydrodynamic instability \cite{LandauHydro} can be also
discussed in terms of the  `Bose condensation'  of the
classical sound or surface waves  
\cite{Zakharov}. 

A second school considers  Bose condensation of 
quasiparticles as a phenomenon of second class, when the
emerging steady state of the system is not in a full thermodynamic
equilibrium,  but is supported by pumping of energy, spin,
atoms, etc. The distribution of quasiparticles in these dynamic states is close
to the thermodynamic equilibrium with a finite chemical potential
which follows from the quasi-conservation of  number of quasiparticles.
In this way, recent experiments in Refs.
\cite{Demokritov} and \cite{Kasprzak}  may be treated as 
Bose condensation of magnons and 
exciton polaritons, respectively (see Appendix G and  also  Ref. \cite{Snoke}; the
possibility of the BEC of quasiequilibrium magnons has been discussed in Ref. \cite{KalafatiSafonov}).
The coherence of these dynamical  states is under investigation \cite{Demidov}.

But not everybody knows that the coherent  spin 
precession discovered  in superfluid $^3$He more than 20 years
ago, and known as  Homogeneously Precessing Domain (HPD),  is
the true Bose-Einstein condensate of magnons (see e.g. reviews 
\cite{Bunkov2007,BunkovVolovik2008}). This spontaneously  emerging steady state
preserves the phase coherence across the whole sample. Moreover,
it is very close to the  thermodynamic equilibrium of the magnon 
Bose condensate and thus exhibits all the superfluid properties
which follow from the off-diagonal long-range order (ODLRO) for
magnons. 

 In the absence of energy pumping this HPD state slowly decays,  
but during the decay the system remains in the state of  the
Bose condensate: the volume of the Bose condensate (the volume
of HPD)  gradually decreases with time without violation of the
observed properties of the spin-superfluid phase-coherent
state. A ste\-a\-dy state of phase-coherent precession can be
supported by pumping. But the pumping need not be coherent -- 
it can be chaotic: the system chooses its own (eigen)
frequency of coherent precession, which emphasizes the
spontaneous emergence of coherence from chaos.

\section{HPD -- magnon Bose condensate}

\subsection{Larmor precession}

The crucial property of the Homogeneously
Precessing Domain is that the Larmor precession 
spontaneously acquires
a coherent phase throughout the whole sample  even in
an inhomogeneous external magnetic field. This is equivalent to
the appearance of a coherent superfluid Bose condensate. It
appears that  the analogy is exact: HPD is the Bose-condensate
of magnons.

The precession of magnetization (spin) occurs after  the
magnetization is deflected by an angle $\beta$ by the rf  field
from its equilibium value
${\bf S}=\chi {\bf H}$ (where  ${\bf H}=H\hat{\bf z}$ is  an
external
 magnetic field and $\chi$ is spin suscepti\-bility of liquid
$^3$He): 
\begin{eqnarray}
 S_x+iS_y =S_\perp e^{i\omega
t+i\alpha}~,
\label{precession}
\\
S_\perp=\chi H\sin\beta~,~S_z=\chi H\cos\beta~.
\label{precession2}
\end{eqnarray}
The immediate analogy \cite{FominLT19} says that in precession the role of the number
density of magnons is played by the deviation of the spin projection
$S_z$ from its equilibrium value:
\begin{equation}
 n_M=\frac{S-S_z}{\hbar}=\frac{S(1-\cos\beta)}{\hbar}~.
\label{NumbaerDensity}
\end{equation}
The number of magnons is a conserved quantity if one neglects
the spin-orbit interaction. It is more convenient to work in
the frame rotating  with the frequency $\omega$ of the rf
field, where the spin is stationary if relaxation is neglected.
The free energy in this frame is
\begin{equation}
F(\beta) =(\omega -\omega_L)S_z +E_{\rm so}(\beta) ~.
\label{FreeEnergy}
\end{equation}
Here  
$\omega_L=\gamma H$ is the Larmor
frequency;  $\gamma$ is the gyromagnetic ratio of the $^3$He atom. The
precession frequency  plays  the role of  the
chemical potential $\mu$ for magnons:
\begin{equation}
 \mu = \omega  ~;
\label{mu}
\end{equation}
while the local Larmor frequency $\omega_L({\bf r})$  plays the role of external potential;
and $E_{\rm so}(\beta)$ is  the energy of spin-orbit
interaction. 
 The properties of the precession  depend on $E_{\rm
so}(\beta)$: stable precession occurs when $E_{\rm
so}(\beta)$ is a concave function of
$\cos\beta$. This is what occurs in $^3$He-B  (see Appendix A).

\subsection{Spectrum of magnons: anisotropic mass}

The important property of the Bose condensation of
magnons in
$^3$He-B is that  the mass of magnons is anisotropic,
i.e. it depends on the direction of propagation  (see Appendix B).  The spectrum of
magnons (transverse spin waves)  is
 \begin{equation}
 \omega({\bf k})= \omega_L  
+\frac{\hbar k_z^2}{2m_M^\parallel(\beta)} +
\frac{\hbar k_\perp^2}{2m_M^\perp(\beta)}~,
\label{MagnonSpectrum}
\end{equation}
where the longitudinal and transverse masses depend on  the
tilting angle. Both masses are much smaller than the mass $m$ of
the $^3$He atom: 
\begin{equation}
\frac{m_M}{m} \sim \frac{\hbar\omega_L}{E_F} ~,
\label{MagnonMass2}
\end{equation}
where $E_F\sim 1K$ is the Fermi energy in
$^3$He liquid  (see Appendix B).
In the co-rotating frame the spectrum is shifted 
 \begin{equation}
 \omega_{\rm co-rot}({\bf k})= \omega_L -\omega
+\frac{\hbar k_z^2}{2m_M^\parallel(\beta)} +
\frac{\hbar k_\perp^2}{2m_M^\perp(\beta)}~,
\label{MagnonSpectrum2}
\end{equation}
which corresponds to the chemical potential $\mu=\omega$.

\subsection{Bose condensation of  magnons}

The value
$\mu=\omega_L$ is critical: when $\mu$ crosses the minimum of the magnon spectrum, the Bose condensation  of magnons with $k=0$  occurs resulting in the  phase-coherent precession of spins and  spin superfluidity at
$\mu>\omega_L$.  
In $^3$He-B, the Bose condensate of magnons is
almost equilibrium. Though the number density of
magnons in precessing state is much smaller than  the density
number $n$ of $^3$He atoms
\begin{equation}
\frac{n_M}{n} \sim  (1-\cos
\beta)\left(\frac{\hbar\omega_L}{E_F}\right)~,
\label{MagnonDensity}
\end{equation}
their mass is also by the same factor smaller (see
Eq.(\ref{MagnonMass2})). The critical temperature
of the Bose condensation of magnons, which follows from this mass is 
\begin{equation}
T_{\rm BEC1}\sim \frac{\hbar^2n_M^{2/3}}{m_M} \sim
\frac{E_F^{4/3}}{(\hbar\omega_L)^{1/3}} \sim 10 E_F~.
\label{CondensationT1}
\end{equation}
The more detailed calculations  gives even higher
 transition temperature (see Appendix C). In any case
the typical temperature of superfluid $^3$He  of order
$T\sim 10^{-3}E_F$  is much smaller than the condensation temperature, and thus the Bose condensation is complete.

\subsection{ODLRO of magnons}

In terms of magnon condensation,  the precession can be viewed
as the off-diagonal long-range order (ODLRO) for magnons. The
ODLRO is obtained  using the Holstein-Primakoff transformation
\begin{eqnarray}
\hat b\sqrt{1-\frac{b^\dagger b}{2S}}= \frac{\hat
S_+}{\sqrt{2S\hbar}}~,
\nonumber
\\
\hat b^\dagger\sqrt{1-\frac{b^\dagger b}{2S}}=
\frac{\hat S_-}{\sqrt{2S\hbar}}~,
\nonumber
\\
\hat b^\dagger\hat b = 
\frac{S-S_z}{\hbar}~.
\label{CreationAnnih}
\end{eqnarray}
In the precessing state of Eq.(\ref{precession}),  the operator
of magnon annihilation has a non-zero vacuum expectation value --
the order parameter:
\begin{equation}
\Psi=\left<\hat b\right>=
\sqrt{\frac{2S}{\hbar}}\sin\frac{\beta}{2}~  e^{i\omega
t+i\alpha}~.
\label{ODLRO}
\end{equation}
So, $\sin(\beta/2)$ plays the role of the modulus of the  order
parameter; the phase of precession $\alpha$ plays the role
of the phase of the superfluid order parameter; and the precession frequency plays the 
role of chemical potential, $\mu=\omega$. Note that for the equilibrium planar ferromagnet, which also can be described in terms of the ODLRO, Eq.(\ref{ODLRO}) does not contain
the chemical potential $\omega$ (see Appendix H); as a result  this analogy with magnon Bose condensation \cite{Nikuni,Giamarchi} (see also \cite{Hohenberg1971}) becomes too far distant.

The precessing angle $\beta$ is typically large in HPD. 
The profile of the spin-orbit energy is such that  at $\mu>\omega_L$ 
(i.e. at $\omega>\omega_L$)  the equilibrium condensate 
corresponds to precession  at  fixed tipping angle (see Appendix A):
\begin{equation}
\cos\beta\approx -\frac{1}{4}~,
\label{EquilibriumTilting}
\end{equation} 
and thus with fixed condensate density:
\begin{equation}
n_M(\mu=\omega_L+0)=|\Psi_{\mu=\omega_L+0}|^2= \frac{5}{4}\frac{S}{\hbar} ~.
\label{InitialCondensate}
\end{equation}

\subsection{Spin supercurrent}

The superfluid mass current carried by magnons is the linear 
momentum of the Bose condensate (see Appendix D):
\begin{equation}
{\bf P}=(S-S_z)\nabla\alpha=  n_M \hbar\nabla\alpha~.
\label{MassCurrentMagnon}
\end{equation}
Here we used the fact that $S-S_z$ and $\alpha$ are canonically 
conjugated variables. This superfluid mass current is
accompanied by the superfluid current of spins transferred by
the magnon condensate. It is determined by the spin to mass
ratio for the magnon, and because the magnon mass is
anisotropic, the spin current transferred by the coherent
spin precession is anisotropic too:
\begin{eqnarray}
J_z=-\frac{\hbar^2}{m_M^\parallel(\beta)}n_M
(\beta)\nabla_z\alpha~,
\label{SpinCurrent1}
\\
{\bf
J}_\perp=-\frac{\hbar^2}{m_M^\perp(\beta)}n_M(\beta)
\nabla_\perp\alpha~.
\label{SpinCurrent2}
\end{eqnarray}
This superfluid current of spins is one more representative of
superfluid currents known or discussed in other systems, such as 
the superfluid current of mass and atoms in superfluid
$^4$He; superfluid current of electric charge in superconductors;
superfluid current of hypercharge in Standard Model; superfluid
baryonic cuurent and current of chiral charge in quark
matter; etc. (recent review on spin currents is in Ref. \cite{SoninReview}). 

This superfluidity  is very similar to superfluidity of the A$_1$ phase of $^3$He where only one spin component is superfluid \cite{VollhardtWolfle}: as a result the superfluid mass current is accompanied by the superfluid spin current.

The anisotropy of the current in Eqs.
(\ref{SpinCurrent1}-\ref{SpinCurrent2}) is an important
modification of the conventional Bose condensation. This effect
is absent in the atomic Bose condensates.

\subsection{Two-domain structure of precession}

The distinguishing property of the Bo\-se condensate of  magnons 
in $^3$He-B is that quasi-equilibrium precession has a fixed  
density of Bose condensate in Eq.(\ref{InitialCondensate}).   
Since the density of magnons in the condensate  cannot relax
continuously, the decay of the condensate can only occur via
the decreasing volume of the condensate.

 \begin{figure}
\centerline{\includegraphics[width=1.0\linewidth]{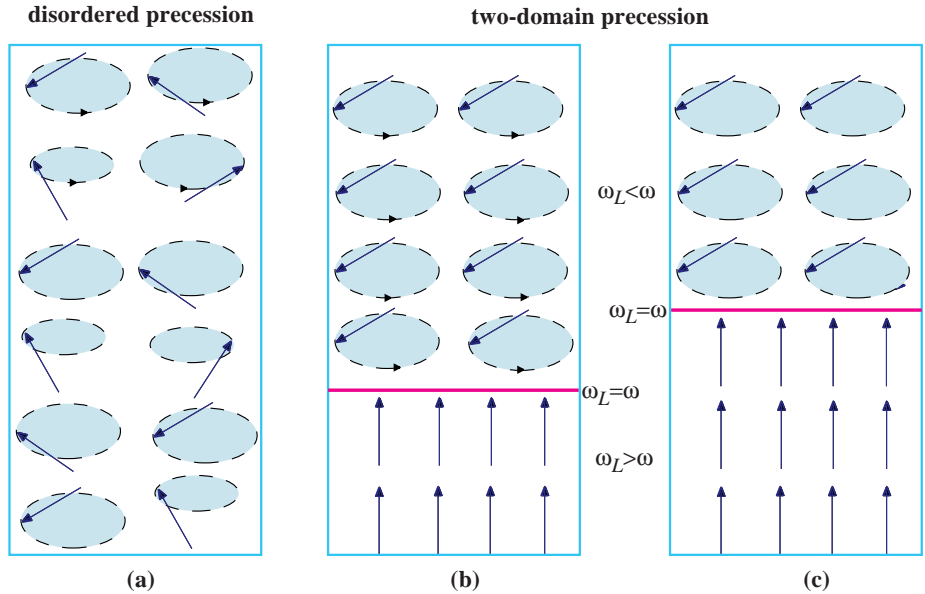}}
  \caption{(a) Disordered precession after pumping of magnons 
spontaneously evolves into  (b) two domains: Homogeneously
Precessing Domain (HPD) with the Bose condensate of magnons, and 
the  domain with static magnetization and no magnons (NPD).  (c)
The  Bose condensate decays due to non-conservation  of the
number of magnons. However, phase coherence is not violated, and
even the density of the Bose condensate $n_M$ does not change.
The relaxation occurs via gradual decrease of the volume of the
Bose condensate.}
  \label{2DomainFig}
\end{figure}

This results in the formation of two-domain precession:  the
domain with the Bose condensate   (HPD) is separated by a phase
boundary from the domain with static equi\-libri\-um ma\-gnetization
(non-precessing do\-main, or NPD). The two-domain structure
spo\-n\-tan\-eously  emerges after the magnetization is deflected
by pulsed NMR, and thus  mag\-nons are pumped into the system 
(Fig. 1(a)). If this happens in an applied gradient of magnetic
field, the  magnons are condens\-ed and collected in the region
of the sample,  where   $\mu\equiv \omega>\omega_L(z)$  (Fig. 1(b)). They
form the HPD there. This process is fully analogous to the
separation of gas and liquid in the gravitational field: the role
of the gravitational field is played by $\nabla \omega_L$. 

In the absence of the rf field, i.e. without continuous pumping,
the precessing domain (HPD) remains in the fully coherent Bose
condensate state,  while the phase boundary between HPD and NPD
slowly moves up so that the volume of the Bose condensate 
gradually decreases (Fig. 1(c)). The frequency $\omega$ of
spontaneous coherence  as well as the phase of precession
remain   homogeneous across the whole Bose condensate domain,
but the magnitude of the frequency changes with time. The
latter is determined by the Larmor frequency at the position of
the phase boundary between HPD and NPD,  
$\mu\equiv\omega= \omega_L(z_0)$;
 in other words at the phase boundary the chemical potential  of
magnons corresponds to the onset of condensation.  

\section{Discussion}

\subsection{Observed superfluid properties of magnon Bose
condensate}

As distinct from the conventional Larmor precession, the phase
coherent precession of magnetizaton in $^3$He-B has all the
properties of the coherent Bose condensate of magnons. The main
spin-superfluid properties of HPD have been
verified already in early experi\-ments 20 years ago
\cite{1,2,3,4,5,6,7}. These include spin supercurrent which
transports the magnetization (analog of the mass current in
conventional superfluids); spin current Josephson effect and
phase-slip processes at the critical current. Later on a spin
current vortex has been observed \cite{Vortex} --  a topological
defect  which is the analog of a quantized vortex in superfluids
and of an Abrikosov vortex in superconductors  (see Appendix D) .

The Goldstone modes of the two-domain structure  of the
Larmor precession have been also observed including the ``sound waves'' of the magnon condensate -- phonos
\cite{5} (see Appendix E) and 
 `gravity' waves - surface waves at the interface between HPD and
NPD \cite{SurfaceWaves}.

\subsection{Exploiting Bose condensate of magnons}

The Bose condensation of magnons in superfluid $^3$He-B has many 
practical applications. 

In Helsinki, owing to the extreme sensitivity  of the Bose
condensate  to textural inhomogeneity, the pheno\-me\-non of Bose
condensation  has been applied to studies of  
supercurrents and topological defects in $^3$He-B. The measurement
technique was called  HPD spectroscopy
\cite{HPDSpectroscopy,HPDSpectroscopy2}. In particular, HPD
spectroscopy provided direct
experimental evidence for broken axial symmetry in the core of
a particular quantized vortex in $^3$He-B. Vortices with
broken symmetry in the core are condensed matter
analogs of the Witten  cosmic strings,  where the
addi\-tional $U(1)$ symmetry is broken inside the string core 
(the so-called superconducting cosmic strings \cite{ewitten}). 
The Goldstone mode of the vortex core resulting from the
spontaneous violation of rotational $U(1)$ symmetry in the core
has been observed
\cite{26}. The so-called spin-mass vortex,  which is a combined
defect serving as the termination line of the topological soliton
wall, has  also been observed and studied using  HPD spectroscopy
\cite{27}. 

In Moscow\cite{Dmitriev}, Grenoble\cite{Grenoble1,Grenoble2}  and
Tokyo\cite{Tokyo1,Sato2008,Tokyo2}, HPD  spectroscopy proved to be extremely
useful for the investigation of the superfluid order parameter in
a novel system -- $^3$He  confined in aerogel.

There are a lot of new physical phenomena related to the  Bose
condensation, which have been observed after the discovery. Other coherently precessing spin states have been observed in $^3$He-B (see review paper \cite{BunkovVolovik2008} and Ref.\cite{Grenoble2}) and also in $^3$He-A \cite{Sato2008}.  These include in particular the compact objects with finite number of the Bose condensed magnons (see  Appendix I). At small number $N$ of the pumped magnons, the system is similar to the Bose condensate of the ultracold atoms in harmonic traps, while at larger $N$ the analog of $Q$-ball in particle physics develops \cite{QBall}.

The important property of the condensation of quasiparticles is that the BEC is the time dependent process. That is why it may experience instabilities which do not occur in the equilibrium condensates of stable particles. In 1989 it was found that the original magnon condensate -- HPD  state -- looses its stability below about 0.4 T$_c$ \cite{CatastrophExp} and experiences catastrophic relaxation. This phenomenon was left unexplained for a long time and only recently the reason was established: in the low-temperature regime, where dissipation becomes sufficiently small, a Suhl instability in the form of spin wave radiation destroys the homogeneous precession \cite{Catastroph}. However,  the magnon BEC in harmonic traps  and $Q$-balls are not destroyed.

In conclusion, the Homogeneously Precessing  Domain discovered
about 20 years ago in superfluid $^3$He-B represented the first
example of a Bose condensate found in nature (if one does not take
into account the strongly interacting superfluid liquid  $^4$He
with its tiny 7\% fraction of the Bose condensed atoms).

  \section{Appendix A. Ginzburg-Landau energy}

  In $^3$He-B  the energy of spin-orbit interaction is
   \begin{eqnarray}
F_{\rm so}(\beta)=\frac{8}{15}\chi\Omega_L^2\left(\cos\beta
+\frac{1}{4}\right)^2 ~~,~~
{\rm when}~~\cos\beta< -\frac{1}{4}~,
 \label{SO1}
   \\
   E_{\rm so}(\beta)=0~~,~~{\rm when}~~\cos\beta> -\frac{1}{4}~.
   \label{SO2}
\end{eqnarray}
Here  $\Omega_L$ is the so-called Leggett frequency; in typical 
NMR experiments  
$\Omega_L^2\ll \omega_L^2$, i.e. the spin-orbit interaction is small compared to Zeeman energy.
 This leads to the following Ginzburg-Landau potential \cite{FominLT19}
 \begin{equation}
F_{\rm so} \left(\vert\Psi\vert\right)=\frac{8}{15}\chi\Omega_L^2 \left(
\frac{\vert\Psi\vert^2}{S}-\frac{5}{4}\right)^2 \Theta \left(
\frac{\vert\Psi\vert^2}{S}-\frac{5}{4}\right)~,
     \label{FHPD}
  \end{equation}
  where $ \Theta(x)$ is Heaviside step function.
 The total Ginzburg-Landau energy functional is 
  \begin{equation}
F_{\rm GL}=\frac{1}{2}\left(m (\vert\Psi\vert)\right)^{-1}_{ij}\nabla_i \Psi^* \nabla_j \Psi+\left(\omega_L({\bf r})- \omega\right) + F_{\rm so} \left(\vert\Psi\vert\right)~,
     \label{TotalGL}
  \end{equation}
  where $m_{ij} (\vert\Psi\vert)$ is the anisotropic mass in the spectrum of magnons, see below Sec. \ref{Sec:MagnonSpectrum}; the local Larmor frequency $\omega_L({\bf r})$ plays the role of the external potential acting on magnons; and the global precession frequency  $\omega$ plays the role of the magnon chemical potential. 
   
   The precession frequency is shifted from the Larmor value when $\cos\beta< -1/4$:
   \begin{equation}
\mu\equiv  \omega= \omega_L   +\frac{4}{15}\frac{\Omega_L^2}{\omega_L}  (1+4\cos\beta)> \omega_L~.
\label{PrecessionFrequency}
\end{equation}
  The Bose condensate starts with
  the tipping angle equal to the so-called Leggett angle, $\cos\beta=-1/4$.
 
  \section{Appendix B. Magnon spectrum}
  \label{Sec:MagnonSpectrum}
  
The spectrum of magnons  in the limit of small spin-orbit interaction is
\begin{equation}
 \omega(k)= \omega_L   +\frac{\hbar k_z^2}{2m_M^\parallel(\beta)} +
\frac{\hbar k_\perp^2}{2m_M^\perp(\beta)}~,
\label{MagnonSpectrum3}
\end{equation}
where the longitudinal and transverse masses depend on the tilting
angle. For superfluid  $^3$He-B  one has the following dependence:  
\begin{eqnarray}
\frac{1}{m_M^\parallel(\beta)}=2
 \frac{c_\parallel^2 \cos\beta+ c_\perp^2(1-\cos\beta)}{\hbar\omega_L} 
,
\nonumber
\\
\frac{1}{m_M^\perp(\beta)}=\frac{c_\parallel^2(1+\cos\beta)
+ c_\perp^2(1-\cos\beta)}{\hbar\omega_L}   ,
\nonumber
\end{eqnarray}
where the parameters $c_\parallel$ and $c_\perp$ are  on  the
order  of the Fermi velocity $v_F$.   In the simplified cases $c_\parallel=c_\perp\equiv c$ , when one
neglects the spin-wave anisotropy (or  in the limit of small tilting
angle), the magnon  spectrum in the limit of small spin-orbit interaction is
\begin{equation}
 \omega(k)= \omega_L   +\frac{\hbar k^2}{2m_M}  ~,
\label{MagnonSpectrumIsotropic}
\end{equation}
where the isotropic magnon  mass is:
  \begin{equation}
m_M =\frac{\hbar\omega_L}{2c^2}~.
\label{IsotropicMass}
\end{equation}
Since $c\sim v_F$,   the relative 
magnitude of the magnon mass compared to the bare mass $m$ of the $^3$He
atom is
  \begin{equation}
\frac{m_M}{m} \sim \frac{\hbar\omega_L}{E_F}~,
\label{MassRatio}
\end{equation} 
where the Fermi energy $E_F\sim mv_F^2\sim p_F^2/m$. 
 
 The spectrum (\ref{MagnonSpectrumIsotropic}) is valid when $kc\ll \omega_L$, however the typical temperature $T$ of HPD is 0.3$T_c$ which is an order of magnitude larger than the magnon gap $\hbar\omega_L \sim 50~\mu K$ at  $\omega_L\sim 1$MHz. That is why one needs the spectrum in the broader range of $k$:
\begin{equation}
 \omega(k)= \pm \frac{\omega_L}{2}   +\sqrt{ \frac{\omega_L^2}{4} +k^2c^2} ~,
 \label{MagnonSpectrumextended}
\end{equation}
where the sign $+$ corresponds to magnons under discussion. Since $T\gg \hbar\omega_L$ thermal magnons are spin waves with linear spectrum $ \omega(k)=ck$, with characteristic momenta $k_T\sim T/\hbar c$. The density of thermal magnons is $n_T \sim k_T^3$.

The density of the condensed of magnons is small when  compared
  to the density of atoms $n=p_F^3/3\pi^2\hbar^3$ in $^3$He liquid
     \begin{equation}
\frac{n_M}{n} \sim \frac{\hbar\omega_L}{E_F} \ll 1~.
\label{DensityRatio}
\end{equation} 
But it is large when compared to the density of thermal magnons:
    \begin{equation}
\frac{n_M}{n_T} \sim \frac{\hbar\omega_L}{T}~\frac{ E_F^2}{T^2} \gg 1~,
\label{DensityToThermalRatio}
\end{equation} 

 \section{Appendix C. Transition temperature}

At first glance, the temperature at which  condensation starts can be estimated as the temperature at which  $n_M$ and $n_T$ are comparable. This gives 
 \begin{equation}
T_{\rm BEC2} \sim \left(\hbar\omega_L  E_F^2\right)^{1/3}~,
\label{CondensationT2}
\end{equation} 
This temperature is much smaller than the estimate in Eq.(\ref{CondensationT1}) which comes from the low-frequency part of the spectrum, but still is much bigger than $T_c$.  

However, since the gap $\omega_L$ in magnon spectrum is small, the condensation may occur even if the number density of the condsensed magnons $n_M\ll n_T$ (see also Refs. \cite{Demokritov,Demidov}). The number of extra magnons which can be absorbed by thermal distribution is  the difference of the distribution function at $\mu=0$ and $\mu\neq 0$. Since $\mu\ll T$, it  is determined by low energy  Rayleigh-Jeans part of the spectrum:
 \begin{equation}
n_{\rm extra}=\sum_{\bf k}\left( \frac{T}{\epsilon -\mu} - \frac{T}{\epsilon} \right)~,
\label{ExtraNumber}
\end{equation} 
Maximum takes place when $\mu=\omega_L$, which gives the dependence of transition temperature on the number of pumped magnons
\begin{equation}
\frac{n_M}{T_{\rm BEC3}}=\frac{1}{4\pi}\int dk~k^2\left( \frac{\omega_L}{k^2c^2} - \frac{\omega_L}{\omega_L^2+k^2c^2}\right)=\frac{\omega_L^2}{4c^3}~.
\label{CondensationT}
\end{equation} 
This gives the transition temperature 
 \begin{equation}
T_{\rm BEC3} =\frac{4n_Mc^3}{\omega_L^2}\sim \frac{E_F^2}{\omega_L}~,
\label{CondensationT3}
\end{equation} 
which is much bigger than the estimations (\ref{CondensationT2}) and (\ref{CondensationT1}).

In any case, at the typical temperatures of superfluid $^3$He  of order
$T\sim 10^{-3}E_F$ the Bose condensation is complete. The hierarchy of temperatures in magnon BEC is thus
 \begin{equation}
\hbar\omega_L \ll T\lll T_{\rm BEC}~.
\label{Temperatures}
\end{equation} 
Bose condensation of magnons in $^3$He-B is similar to the Bose condensation of ultrarelativistic particles with spectrum $E(p) =\sqrt{M^2c^4 + c^2p^2}$ in the regime when 
 \begin{equation}
 Mc^2\ll T\ll T_{\rm BEC} 
~.
\label{URgas}
\end{equation} 

For the Bose gas in laser traps, the hierarchy of temperatures is $\hbar\omega_{\rm h} \ll T<T_{\rm BEC}$, where  $\omega_{\rm h}$ is the frequency in the harmonic trap \cite{Pit1999}.

      \section{Appendix D. Superflow}
 
In the simple case of isotropic mass the kinetic energy of superflow of magnon BEC in the London limit is
\begin{equation}
E_{\rm kin}=\frac{1}{2} \rho_{sM} {\bf v}_{sM}^2~~,~~
 {\bf v}_{sM} =\frac{\hbar}{m_M} \nabla\alpha~~,~~ \rho_{sM} =n_M m_M~.
\label{KineticEnergy}
\end{equation} 
Here $ {\bf v}_{sM}$ and $\rho_{sM}$ are superfluid velocity and density of magnon superfluidity.  Note that  the total number of magnons enters  $\rho_{sM}$, since the temperature is low, and the number
of thermal magnons is negligibly small.
In the Ginzburg-Landau regime this has the form:
\begin{equation}
E_{\rm kin}=\frac{\hbar^2}{2m_M} \vert\nabla\Psi\vert^2~,
\label{KineticEnergyGL}
\end{equation} 
and the total GL functional is
\begin{equation}
E_{\rm GL}=\frac{\hbar^2}{2m_M} \vert\nabla\Psi\vert^2 + (\omega_L-\mu)\vert\Psi\vert^2+F_{\rm so}(\Psi)~,
\label{EnergyGL}
\end{equation} 

The mass supercurrent generated by magnons is
\begin{equation}
{\bf P}=\rho_{sM} {\bf v}_{sM}  =\hbar n_M \nabla\alpha~,
\label{MassCurrent}
\end{equation} 
which gives Eq.(\ref{MassCurrentMagnon}).

Circulation quantum is
\begin{equation}
\kappa_M=\oint d{\bf r}\cdot  {\bf v}_{sM}  =\frac{2\pi\hbar}{m_M}~.
\label{Circulation}
\end{equation} 
This demonstrates that the observed  spin vortex \cite{Vortex} with nonzero winding of $\alpha$ has also 
circulation of the mass current. This is similar to the A$_1$ phase of $^3$He with the superfluidity of only one spin component: the superfluid mass current is accompanied by the superfluid spin current.

  \section{Appendix E. Phonons in magnon superfluid and symmetry breaking field}

The speed of sound in magnon superfluid  is determined by compressibility of magnons gas, which is non-zero due to dipole interaction. Applying Eq.(\ref{EnergyGL}) in the HPD region, i.e. in the  region where $\cos\beta< -1/4$, and using the isotropic spin wave approximation $c_\parallel=c_\perp$ 
(for anisotropic spin wave velocity see e.g. Ref. \cite{VolovikPhonons}) one obtains  the sound with velocity:
\begin{equation}
c_s^2= \frac{1}{m_M}\frac{dP}{dn_M}=   \frac{n_M}{m_M}\frac{d\mu}{dn_M}=  \frac{n_M}{m_M}\frac{d^2E_{\rm so}}{dn_M^2} ~.
\label{SpeedSound}
\end{equation}
For $\cos\beta$ close to $-1/4$ the speed of sound is
\begin{equation}
c_s^2= \frac{8}{3}\frac{\Omega_L^2}{\omega_L^2}c^2~.
\label{SpeedSound2}
\end{equation}
This sound observed in Ref. \cite{5} is the Goldstone mode of the magnon Bose condensation.
 
 The important property of magnon BEC is that the Goldstone boson (phonon) acquires mass (gap) due to the transverse RF field ${\bf H}_{\rm RF}$. The latter plays the role of the symmetry breaking field, since it violates the $U(1)$ symmetry of precession \cite{VolovikPhonons}. As a result, the extra term in the Ginzburg-Landau energy, $F_{\rm RF} = -\gamma {\bf H}_{\rm RF}\cdot{\bf S}$, induced by ${\bf H}_{\rm RF}$ depends explicitly on the phase of precession $\alpha$ with respect to the direction of the RF-field in rotating frame. For $\cos\beta =-1/4$ one has
\begin{equation}
F_{\rm RF}= -  \gamma H_{\rm RF} S_\perp \cos\alpha  \approx - \frac{\sqrt{15}}{4} \left(1-\frac{\alpha^2}{2}  \right) \gamma H_{\rm RF} S~.
\label{SymmetryBreakingRF}
\end{equation}
This term adds the mass (gap) to the phonon spectrum:
\begin{equation}
\omega^2_s(k)= c_s^2k^2  + m_s^2~~,~~ m_s^2=\frac{4}{\sqrt{15}} \gamma H_{\rm RF} \frac{\Omega_L^2}{\omega_L}~.
\label{SoundMode}
\end{equation}
The gap $m_s$ of the Goldstone mode induced by the symmetry breaking field has been measured in Refs. \cite{PhononMass,Skyba}.

  \section{Appendix F. Critical velocities and vortex core}

The speed of sound in magnon gas determines the Landau critical velocity  of the counterflow at which phonons are created:
\begin{equation}
v_{\rm L}=c_s~.
\label{LandauVelocity}
\end{equation}
For conventional condensates this suggests that the coherence length and the size of the vortex core should be:
\begin{equation}
  \frac{\hbar}{m_M c_s} \sim \frac{c}{\Omega_L} \sim \xi_D~.
\label{CoreRadius}
\end{equation}

 \begin{figure}
\centerline{\includegraphics[width=1.0\linewidth]{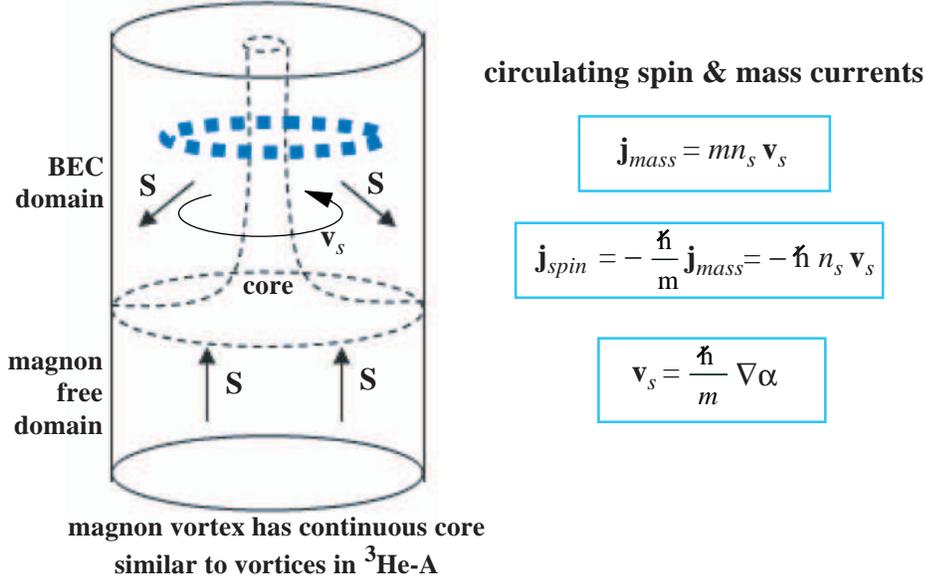}}
  \caption{Spin-mass vortex in magnon BEC. The spin current around the vortex core is accompanied 
  by mass current.  According to Eq.(\ref{CoreRadius}) the size of the vortex core diverges when the HPD domain boundary is approached where the local Larmor frequency $\omega_L=\omega$.}
  \label{VortexFig}
\end{figure}

However, for magnons BEC in $^3$He-B this gives only the lower bound on the core size. The core is larger due to specific profile of the Ginzburg-Landau (dipole) energy in Eq.(\ref{FHPD}) which is strictly zero for $\cos\beta>-1/4$. This leads to the special topological properties of coherent precession (see  Ref. \cite{MisirpashaevVolovik1992}). As a result the spin vortex created and observed  in Ref. \cite{Vortex}  has a continuous core with broken symmetry, similar to vortices in superfluid $^3$He-A \cite{SalomaaVolovik1987}. The size of the continuous core is determined by the proper coherence length   \cite{Fomin1987} which can be found from the competition between the first two terms in Eq.(\ref{EnergyGL}):
\begin{equation}
r_{\rm core}\sim   \frac{\hbar}{\sqrt{m_M(\mu -\omega_L)} }  \sim   \frac{c}{\sqrt{\omega_L(\omega -\omega_L)}}   ~.
\label{CoreRadiusLarge}
\end{equation}  
This coherence length determines also the critical velocity for creation of vortices:
\begin{equation}
v_c\sim \frac{\hbar}{m_M r_{\rm core}}~.
\label{CritVelocity}
\end{equation}

For  large tipping angles of precession  the symmetry of the vortex core is restored: the vortex becomes singular with the core radius
$r_{\rm core}\sim \xi_D$ \cite{SoninVortex}.

   \section{Appendix G. BEC in YIG}

Magnons in yttrium-iron garnet (YIG) films have the quasi 2D spectrum:
\cite{Demokritov,Demidov} 
\begin{equation}
\omega_n(k_x,k_y)= \Delta_n + \frac{k_y^2}{2m_y} + \frac{(k_x-k_0)^2}{2m_x}  ~,
\label{2Dspectrum}
\end{equation} 
where magnetic field is along $x$; the gap in the lowest branch $\Delta_0=2.1$ GHz $\equiv 101$ mK at $H=700$ Oe \cite{Demokritov} and $\Delta_0=2.9$GHz at $H=1000$ Oe \cite{Demidov}; the position of the minimum $k_0=5\cdot 10^4$ 1/cm \cite{Demokritov}; the anisotropic magnon mass can be probably estimated as 
$m_x\sim k_0^2/\Delta$ with $m_y$ being somewhat bigger, both are of order of electron mass. 

If one neglects the contribution of the higher levels and consider the 2D gas, the Eq.(\ref{ExtraNumber}) becomes
 \begin{equation}
n_{\rm extra}=  \frac{T}{2\pi \hbar} \sqrt{m_x m_y}~ \ln \frac{\Delta_0}{\Delta_0 -\mu}~,
\label{ExtraNumber2D}
\end{equation} 
In 2D, all extra magnons can be absorbed by thermal distribution at any temperature without formation of Bose condensate. The larger is the number  $n_M$ of the pumped magnons the closer is $\mu$ to $\Delta_0$, but  $\mu$ never crosses $\Delta_0$. At large $n_M$ the chemical potential 
exponentially approaches  $\Delta_0$ from below and the width of the distribution becomes exponentially narrow:
\begin{equation}
  \frac{(\delta k_y)^2}{m_y \Delta_0} \sim  \frac{(\delta k_x)^2}{m_x\Delta_0} \sim \frac{\Delta_0-\mu}{\Delta_0}\sim\exp\left( -\frac{2\pi \hbar n_M }{T \sqrt{m_x m_y}}\right) ~.
\label{Width}
\end{equation} 
If one uses the 2D number density  $n_M=\delta N ~d$ with the film thickness $d=5~\mu$m and 3D number density $\delta N\sim 5\cdot 10^{18}$~cm$^{-3}$, one obtains that at room temperature the exponent is
\begin{equation}
\frac{2\pi \hbar n_M }{T \sqrt{m_x m_y}}\sim  \frac{2\pi  \Delta_0 }{T}   \frac{\delta N~d }{ 10 k_0^2}  \sim  10^2 ~.
\label{Exponent}
\end{equation} 
If this estimation is correct, the peak should be extremely narrow, so that all extra magnons are concentrated at the lowest level of the discrete spectrum. However, there are other contributions to the width of the peak due to:  finite resolution of spectrometer, magnon interaction,  finite life time of magnons and the influence of the higher discrete levels $n\neq 0$.

In any case, the process of the concentration of extra magnons in the states very close to the lowest energy is the signature of the BEC of magnons.  The main property of the  room temperature  BEC in YIG  is that the transition temperature $T_c$ is only slightly higher than temperature, $T_c-T\ll T$; as a result the 
number of  condensed magnons  is small compared to the number of thermal magnons: $n_M\ll n_T$.  Situation with  magnon BEC in $^3$He is the opposite,  one has $T\ll T_c$ and thus $n_M\gg n_T$.

 \section{Appendix H. Magnon BEC vs planar ferromagnet}
 
Coherently precessing state HPD state
in $^3$He-B has
\begin{equation}
 S_x+iS_y =S_\perp e^{i\omega t+i\alpha}~.
\label{ODLRO1}
\end{equation}
The coherent  state in YIG has
\begin{equation}
 S_x+iS_y =S_\perp \cos(k_0x)e^{i\omega t+i\alpha}~.
\label{ODLROYIG}
\end{equation}
For equilibrium planar ferromagnet one has \cite{Hohenberg1971,Nikuni}
\begin{equation}
 S_x+iS_y =S_\perp e^{i\alpha}~.
\label{ODLRO2}
\end{equation}
This means that the equilibrium planar ferromagnet can be also described in terms of the ODLRO.

Magnons were originally determined as second quantized modes in the background of stationary state with magnetization along $z$ axis. Both the static state in Eq.(\ref{ODLRO2}) and precessing states (\ref{ODLRO1}) and (\ref{ODLROYIG}) can be interpreted as BEC of these original magnons.
On the other hand, the stationary and precessing states can be presented as new vacuum states,  the time independent and the time dependent vacua respectively. The excitations --  phonons -- are the  second quantized modes in the background of  a new vacuum.
What is the principle difference between the stationary vacuum of planar ferromagnet and the time dependent vacuum of coherent precession?

The major point which distinguishes the HPD state (\ref{ODLRO1}) in $^3$He-B and the  coherent precession (\ref{ODLROYIG}) in YIG from the equilibrium magnetic states is the conservation (or quasi-conservation) of the $U(1)$ charge $Q$.  The charge $Q$ is played by the spin projection $S_z$ in magnetic materials  (or the related number $N$ of magnons) and by number $N$ of atoms in atomic BEC.  This conservation gives rise to the chemical potential $\mu=dE/dN$, which is the precession  frequency $\omega$ in magnetic systems. On the contrary, the static state in Eq.(\ref{ODLRO2}) does not contain the chemical potential $\omega$, i.e.   the conservation   is not  in the origin of formation of the static equilibrium state; the chemical potential of magnons in a fully equilibrium state is always strictly zero, $\mu=\omega=0$. 

The spin-orbit interaction violates the conservation of $S_z$, as a result the life time of magnons is finite.  For the precessing states (\ref{ODLRO1}) and (\ref{ODLROYIG}) this leads to the finite life time of the coherent precession.  To support the steady state of precession the pumping of spin and energy is required. On the contrary,  the spin-orbit interaction does not destroy the long-range magnetic order in the static state (\ref{ODLRO2}): this is fully  equilibrium state which does not decay and thus does not require pumping: the life time of static state is macroscopically large and thus by many orders of magnitude exceeds the magnon relaxation time. That is why a planar ferromagnet (\ref{ODLRO2}) is just one more equilibrium state of quantum vacuum, in addition to the easy axis ferromagnetic state, rather than the magnon condensate.

The property of (quasi)conservation of the $U(1)$ charge $Q$  distinguishes the coherent precession from the other coherent phenomena, such as optical lasers and standing waves.
For the real BEC one needs the conservation of particle number or charge $Q$
during the time of equilibration. BEC occurs due to the thermodynamics, when the number of particles
(or charge $Q$) cannot be 
accommodated by thermal distribution, and as a result the extra part must be accumulated 
in the lowest energy state. This is the essence of BEC.

Photons and phonons can also form the true BEC (in thermodynamic sense) under pumping, 
again if the lifetime is larger than thermalization time.  These BEC states are certainly 
different from such coherent states as optical lasers and from the equilibrium deformations of solids.

 \section{Appendix I. Finite-size BEC  \& $Q$-ball}

 When $\cos\beta >-1/4$ ($\beta< 104^\circ$),
the
 spin-orbit   interaction $F_{\rm so}(\Psi)$  becomes nonzero  at finite polar angle $\beta_L$ of vector ${\bf L}$ of $^3$He-B orbital angular momentum:
 \begin{equation}
 F_{\rm so}(\Psi)=\chi\Omega_L^2\left[   \frac {4\sin^2(\beta_L/2)}{5S}
\vert\Psi\vert^2- \frac {\sin^4(\beta_L/2)}{S^2} \vert\Psi\vert^4
\right],
     \label{FD}
  \end{equation}
As a result  the texture of vector ${\bf L}$  forms the  potential well for magnons.

Four  regimes of magnon condensation are possible in the potential well, which  correspond to four successive ranges of the $U(1)$ charge $Q$ (magnon number $N$). (i) At the smallest $N$ the interaction can be neglected and the non-interacting magnons occupy the lowest energy state in the potential  well. (ii) With increasing $N$ the Thomas-Fermi regime of interacting magnons is reached. 
(iii) When the number of magnons is sufficiently large, they start to modify the potential well; this is the regime of the so-called $Q$-ball \cite{QBall}.  (iv)
Finally when the size of the $Q$-ball reaches the dimension of the experimental cell the homogeneous BEC is formed -- the HPD.

The first two regimes are similar to what occurs in atomic BEC in laser traps, though not identical.
The difference comes from the 4-th order term in the GL energy (\ref{FD}) which demonstrates that  the attractive interaction between magnons is determined by the texture
 \begin{equation}
 F_{\rm so}(\Psi)=U(r)\vert\Psi\vert^2 +V(r)\vert\Psi\vert^4~,
     \label{Potentials}
  \end{equation}
In the simplest case of the spherically symmetric  harmonic trap one has  
 \begin{equation}
  U(r) =m_M\omega_{\rm h}^2r^2~~,~~V(r)=-\mu' \left(\frac{\omega_h r}{c_s}\right)^4~,
       \label{Harmonic}
  \end{equation}
where $\omega_{\rm h}$ is the harmonic oscillator frequency;  the magnon interaction is normalized  to its magnitude $\mu'=d\mu/dn_M$ in the HPD state and to the speed of sound  $c_s$ also in the HPD state.

The first two regimes can be qualitatively described using simple dimensional analysis. Let $r_N$ is the dimension of magnon gas as a function of the magnon number $N$. Taking into account that $N\sim \vert \Psi\vert^2 r_N^3$  one estimates the GL energy (\ref{EnergyGL}) of the condensate as:
 \begin{equation}
F\sim N \left(\omega_L +\frac{3}{4}\omega_{\rm h}\left(\frac{r_{\rm h}^2}{r_N^2} +
\frac{r_N^2} {r_{\rm h}^2}\right)  -\gamma N r_N\right)~.
\label{DimensionalAn}
\end{equation}
Here $r_{\rm h}=(m_M \omega_{\rm h})^{-1/2}$ is the harmonic oscillator length; the prefactor $3/4$ is introduced to match the oscillator frequency after minimization over $r_N$ in the linear regime;  and $\gamma =\mu' \left(\omega_h /c_s\right)^4$. Minimization at fixed magnon number $N$ gives two regimes. 

(i) In the regime linear in $N$ (the regime of spin-waves) one obtains the $N$-independent radius $r_N=r_{\rm h}$; and
 the Bose condensation occurs at the lowest energy level which corresponds to the precession with frequency (chemical potential)
\begin{equation}
 \omega-\omega_L  \approx\frac{3}{2}\omega_{\rm h}~~,~~N\ll N_1=\frac{\omega_{\rm h}}{ r_{\rm h}\gamma}~.
\label{SpinWaveRegime}
\end{equation}

(ii) At larger $N\gg N_1$ one obtains the analog of the Thomas-Fermi droplet whose size
and  precession frequency depend on $N$:
\begin{equation}
r_N\sim ~r_{\rm h} \frac{N}{N_1}~~,~~ \omega -\omega_L\sim  - \omega_{\rm h}\frac{N^2}{N_1^2}~~,~~N\gg N_1~.
\label{NonlinearRegimeRadius}
\end{equation}
For comparison,  the atomic BEC in the nonlinear regime, which can be obtained by the same procedure, is characterized by \cite{Pit1999}
\begin{equation}
r_N\sim ~r_{\rm h} \left(\frac{N}{N_1}\right)^{1/5}~~,~~
\mu \sim \omega_{\rm h} \left(\frac{N}{N_1}\right)^{2/5}~~,~~N\gg N_1~,
\label{NonlinearRegimeBEC}
\end{equation}
where $N_1=r_{\rm h}/a$ with $a$ being the scattering length.
 
 (iii) The  regime of $Q$-ball emerges when the density of magnons in center of the droplet is such that $\beta$ is not small, and the spin-orbit interaction starts to modify the potential well. This regime develops when $N$ approaches the characteristic value $N_2$:
 \begin{equation}
N_2=N_1~ \frac{c_s}{\omega_{\rm h}r_{\rm h}}\sim  \frac{c_s}{r_{\rm h}^2\gamma}
\sim N_1~ \sqrt{\frac{m_Mc_s^2}{\omega_{\rm h}}}~.
\label{N_2}
\end{equation}

(iv) Finally, the HPD emerges when the size of the droplet reaches the size $L$ of the cell, and the magnon number $N$ reaches the maximum value $N_{\rm max}\sim n_ML^3$, where $n_M$ is magnon density 
in HPD.

A typical texture in the vessel is determined by the vessel geometry and thus by dimension $L$ of the cell, $\beta_L\sim r/L$. In such cases, the corresponding parameters are
\begin{equation}
\omega_{\rm h}\sim \frac{c_s}{L}~,~r_{\rm h}\sim \sqrt{L\xi_D}~, ~N_1\sim N_{\rm max}\left(\frac{\xi_D}{L}\right)^{1/2},~N_2\sim N_{\rm max}~.
\label{TypicalValues}
\end{equation}
Here $\xi_D=c/\Omega_L \sim 10^{-3}$ cm is the dipole length. Since $N_2\sim N_{\rm max}$, the $Q$-ball regime develops at $N\lesssim N_{\rm max}$, and then $Q$-ball  transforms to HPD when $N$ approaches $N_{\rm max}$. In the applied external gradient of magnetic field  the two-domain state in Fig. \ref{2DomainFig} emerges earlier.

One may expect a similar finite size BEC of magnons in rotating $^3$He-A, where the trap for magnons is provided by the core of individual continuous $4\pi$ vortex
\cite{SalomaaVolovik1987}. Applying gradient of magnetic field one may study individual vortices by the magnon BEC tomography.

\begin{acknowledgements}
I am grateful to Yu.M. Bunkov,  S. Demokritov, V.V. Dmitriev,  V.B. Eltsov, I.A. Fomin, M. Krusius and V.S. L'vov  for 
illuminating  discussions. This work was supported in part by 
ESF COSLAB Programme and by the Russian Foundation for
Basic Research (grant 06-02-16002-a).
\end{acknowledgements}

   
\end{document}